\documentclass{aa}  
\usepackage{graphicx}
\usepackage{placeins}
\usepackage{txfonts}
\usepackage{lipsum}
\usepackage{float}
\usepackage{subcaption}         % necessary for continued figures, example in section 3
\usepackage[utf8]{inputenc}
\usepackage{newunicodechar}
\newunicodechar{≈}{\approx}
\usepackage{lscape}             % to rotate a single page table, example in appendix.
\usepackage{placeins}           % useful with \FloatBarrier, to keep 
\usepackage[colorlinks=true,
            linkcolor=blue,
            citecolor=blue,
            urlcolor=blue]{hyperref}
\usepackage{xcolor}
           
\begin{document}
\titlerunning{What flares in galactic nuclei?}
\authorrunning{Eduardo Concepci\'on et al.}
   \title{The ePESSTO+ census of ZTF nuclear transients. I. What flares in galactic nuclei?}

%   \subtitle{Subtitle}

%%%%%%%%%%%%%%%%%%%%%%%%%%%%%%%%%%%%%%%%
% Please do not include ORCIDs next to author names.
% Only ORCIDs authenticated by individual authors in EDP Sciences editorial system will be taken into account.
% ORCIDs included here will be removed.
%%%%%%%%%%%%%%%%%%%%%%%%%%%%%%%%%%%%%%%%

\author{E.~Concepci\'on\inst{1}\corrauth{eduardo.concepcion@ung.si}
  \and T.~Petrushevska\inst{1}
  \and G.~Leloudas\inst{\ref{DTU}}
    \and D. Ili\'{c}\inst{\ref{Belgrade}}
    \and M. Gromadzki\inst{\ref{Warsaw}}
    \and S. Bhattacharyya \inst{1}
    \and P. Charalampopoulos \inst{\ref{Turku}, \ref{ICE}}
    \and A. Gomboc\inst{1}
    \and N. Ihanec \inst{\ref{INGT}}
    \and M. Nicholl\inst{\ref{QUB}}
    \and P. J. Pessi \inst{\ref{NCBJ}}
    \and D. R. Young \inst{\ref{QUB}}
    \and P. Wiseman \inst{\ref{Southampton}}
    \and T.-W. Chen \inst{\ref{Taiwan}}
    \and M. Dennefeld \inst{\ref{Paris}}
    \and Y. Dgany\inst{\ref{Tel Aviv}}
    \and C. P. Guti\'errez \inst{\ref{ICE}, \ref{IEEC}}
    \and C. Inserra \inst{\ref{Cardiff}}
    \and L. Makrygianni\inst{\ref{Lancaster}}
    \and T. E. Müller-Bravo \inst{\ref{Dublin}, \ref{A. Prat}}
    \and S. Nissen \inst{\ref{DTU}}
    \and F. Onori\inst{\ref{INAF}}
    \and R. Roy\inst{\ref{KLEF}}
    \and M. Quilt \inst{\ref{Southampton}}}
 % \and More\inst{}}

\institute{%
  Center for Astrophysics and Cosmology, University of Nova Gorica, 
  Vipavska 11c, 5270 Ajdovščina, Slovenia. 
  \label{CAC}
  \and
  DTU Space, National Space Institute, Technical University of Denmark, 
  Elektrovej 327, 2800 Kgs. Lyngby, Denmark.
  \label{DTU}
  \and
  Department of Astronomy, University of Belgrade -- Faculty of Mathematics, 
  Studentski trg 16, 11000 Belgrade, Serbia.
  \label{Belgrade}
   \and
  Astronomical Observatory, University of Warsaw, Al. Ujazdowskie 4, 00-478 Warszawa, Poland
  \label{Warsaw}
  \and 
 Finnish Centre for Astronomy with ESO (FINCA), FI-20014 University of Turku, Finland
  \label{Turku}
  \and 
  Institute of Space Sciences (ICE, CSIC), Campus UAB, Carrer de Can Magrans, s/n, E-08193 Barcelona, Spain
  \label{ICE}
  \and 
    Isaac Newton Group of Telescopes, Apartado de correos 321, E-38700, Santa Cruz de La Palma, Canary Islands; Spain 
  \label{INGT}
    \and 
    Astrophysics Research Centre, School of Mathematics and Physics, Queens University Belfast, Belfast BT7 1NN, UK
  \label{QUB}
   \and 
  Astrophysics Division, National Centre for Nuclear Research, Pasteura 7, 02-093 Warsaw, Poland
  \label{NCBJ}
  \and 
  School of Physics and Astronomy, University of Southampton, Southampton, SO17 1BJ, UK
  \label{Southampton}
   \and 
  Graduate Institute of Astronomy, National Central University, 300 Jhongda Road, 32001 Jhongli, Taiwan
  \label{Taiwan}
  \and 
  IAP/CNRS and Sorbonne University, Paris
  \label{Paris}
  \and School of Physics and Astronomy, Tel Aviv University, Tel Aviv 69978, Israel
  \label{Tel Aviv}
  \and 
  Institut d'Estudis Espacials de Catalunya (IEEC), 08860 Castelldefels (Barcelona), Spain
  \label{IEEC}
  \and 
  Cardiff Hub for Astrophysics Research and Technology, Department of Physics, School of Physical, Chemical and Environmental Sciences, Cardiff University, Queens Buildings, The Parade, Cardiff, CF24 3AA, UK
  \label{Cardiff}
  \and
  School of Physics and Astronomy, Lancaster University, Lancaster LA1 4YB, UK.
  \label{Lancaster}
\and 
  School of Physics, Trinity College Dublin, The University of Dublin, Dublin 2, Ireland
  \label{Dublin}
  \and 
  Instituto de Ciencias Exactas y Naturales (ICEN), Universidad Arturo Prat, Chile
  \label{A. Prat}
    \and 
   Department of Physics, Koneru Lakshmaiah Education Foundation (KLEF), Green Fields, Vaddeswaram, Guntur District, Andhra Pradesh, Pincode: 522302, India.
  \label{KLEF}
   \and
  INAF -- Osservatorio Astronomico di Roma, 
  Via Frascati 33, 00078 Monte Porzio Catone (RM), Italy.
  \label{INAF}}

   \date{\today}

% \abstract{}{}{}{}{}
% 5 {} token are mandatory
   \abstract
% context
{Luminous transients in galactic nuclei arise from several distinct processes,
only some of which trace the supermassive black hole (SMBH). These include
tidal disruption events (TDEs), changing-look active galactic nuclei, Bowen
fluorescence flares (BFFs) and ambiguous nuclear transients, alongside
supernovae that appear nuclear but are unrelated to the SMBH. Reliable
classification currently requires spectroscopy, which will not scale to the
alert volumes expected from the Legacy Survey of Space and Time (LSST).}
% aims
{We aim to characterize the population of nuclear transients accessible to
real-time photometric selection, to identify early-time observables capable of
distinguishing SMBH-related flares, and TDEs in particular, from the dominant
supernova background, and to assess the biases this selection introduces.}
% methods
{We implemented a real-time filter on the Lasair broker, selecting transients
from the Zwicky Transient Facility (ZTF) alert stream by nuclear coincidence
with the host galaxy alone, with no cuts on colour, light-curve shape or prior
variability, so as to retain SMBH-related flares in both quiescent and active
hosts. The filter ran over 16 months, with candidates reported for
classification through ePESSTO+ and other facilities.}
% results
{The filter returned 476 candidates, of which 202 were spectroscopically
classified, yielding predominantly supernovae, 10 confirmed TDEs ($5\%$) and one
BFF. The mean $g-r$ colour, measurable before peak, and the rise time separate
TDEs from the supernova background. Our filter recovers 10 of the 25 TDEs
classified in the same period and declination range; most misses are technical
or observational rather than physical.}
% conclusions
{Simple real-time filtering combined with basic photometric criteria can already isolate TDEs efficiently from nuclear transient alert streams, and high completeness is achievable. In the LSST era, where samples will be predominantly photometric, selection bias will become as important as contamination, and retaining flares in active hosts is necessary if unusual events are not to be excluded.}

    
   \maketitle
\nolinenumbers

%%%%%%%%%%%%%%%%%%%%%%%%%%%%%%%%%%%%%%%%%%%%%%%%%%%%%%%%%%%%%%

%\clearpage
\section{Introduction}
What flares in the nucleus of a galaxy? The question is deceptively simple: a luminous transient coincident with a galactic center can arise from several physically unrelated processes, only some of which trace the supermassive black hole (SMBH) which is found there. When they do, these flares offer powerful probes of the extreme environments surrounding SMBHs  \citep[e.g.][]{EHT2019_M87_I}, of their accretion processes, and of their galactic environments. The SMBH-related transient population is itself heterogeneous, encompassing tidal disruption events (TDEs), changing-look active galactic nuclei \citep[CLAGN;][]{Camus2026}, Bowen fluorescence flares  \citep[BFF;][]{BFF}, and a growing number of ambiguous nuclear transients (ANTs) whose nature is unresolved \citep{Wiseman2025, 2025SciA...11...74H} . Along these sit events unrelated to SMBHs that nonetheless appear nuclear, mainly supernovae (SNe) occurring in dense nuclear clusters or projected along the line of sight. In most cases, the information obtained from the transients' lightcurve is limited by survey-specific cadences and bands, and spectroscopic observations are needed for reliable classification. Even then, some photometric and spectroscopic features can be common to different types of transients, which further complicates the general characterization of the nuclear transient population. Although luminous transients in galactic centres are routinely identified by wide-field optical surveys such as the Zwicky Transient Facility \citep[ZTF;][]{Graham2019, ZTF_Bellm, ZTF_Masci, ZTF_Dekany} or the Asteroid Terrestrial-impact Last Alert System \citep[ATLAS;][]{atlas}, Vera C. Rubin Observatory's Legacy Survey of Space and Time \citep[LSST;][]{LSST} is expected to discover substantially larger numbers of these events, as predicted by simulations of nuclear transient detection rates \citep{Bricman}. This makes spectroscopic classification for all events unfeasible. It is therefore critical to understand the statistics of the nuclear transient population as a whole and to develop efficient photometric strategies for selecting SMBH-related flares, so that events tracing SMBH activity, or exhibiting unusual behaviour, can be identified early and follow-up resources allocated accordingly.
	
	%TDEs
Of the SMBH-related transients, TDEs have attracted particular effort because the flare itself encodes the mass of the disrupting SMBH \citep[see e.g.][]{2019ApJ...872..151M, Ryu_2020, Sarin_2024, mummery2024opticaluvplateauxraytidal}. Searches have accordingly concentrated on quiescent hosts, where a flare stands out against a passive background, but that boundary is not clean: a few TDE candidates have also been reported in active galaxies, and there is growing evidence for repeating flares in active nuclei, with TDEs being one of the possible mechanisms behind them \citep[e.g.][]{Petrushevska,Dennefeld}. TDEs occur when a star that passes close enough to an SMBH is disrupted by its tidal forces, and circularization and subsequent accretion of part or all of the stellar debris can produce a bright flare \citep{Hills, Rees}.  The current sample of spectroscopically confirmed TDEs on the Transient Name Server (TNS)\footnote{https://www.wis-tns.org/} comprises  $\sim150$ events \citep[e.g.][]{Langis_2026}, which reflects their  intrinsically low occurrence rate of $3.2 \times 10^{-5}$ per galaxy per year \citep{TDE_rate}, roughly two orders of magnitude below the rate of Type Ia supernovae in a comparable galaxy \citep[e.g.][]{Sn_rate, SN_rates}. Consequently, TDEs constitute on the order of 1\% of all spectroscopically classified transients on the TNS. This is, however, also biased by the small number of groups that actively classify nuclear transients, as highlighted by the identification of several non-spectroscopically confirmed TDEs from archival data \citep{Lost_and_found} that were missed in real-time. 

Observationally, optical lightcurves of TDEs typically exhibit a rise on the order of weeks to a few months followed by a gradual, slower decline, resembling those of supernovae. The peak absolute $r$-band magnitude of these events sits in a range between $-18$ and $-21$ \citep[e.g.][]{Hammerstein}, with a faint subclass reaching only about $-17$ \citep{blagorodnova,faint_TDEs}. Rare, exceptionally luminous events such as ASASSN-15lh reached $M \approx -23.5$ in the $g$ band \citep{assasn15_TDE}. The post-peak decay of a full disruption is expected to follow a $t^{-5/3}$ power law, corresponding to the fallback rate of the stellar debris \citep{Rees1990}, although this is not an empirical value and not all events follow this relation strictly \citep[e.g.][]{Lodato2009, Lodato2011}. Additionally, partial or repeating TDEs have also been observed, where a star is only partially disrupted over multiple pericenter passages \citep{partial}, resulting in more than one flare at the same location. TDEs usually maintain a consistently blue colour with little evolution \citep{Seventeen}, which in combination with their nuclear location can be used to select potential candidates. Common spectral features in TDEs include broad H and He lines \citep{Arcavi_14}, although these are not found in all TDEs. In some cases, broad Bowen emission lines \ion{O}{iii}~$\lambda3760$ and \ion{N}{iii}~$\lambda4100,\lambda4640$ are sometimes also found in their spectra \citep{TDE_bowen, Leloudas_19}. However, not all of these features are present in all events. As a result, several spectroscopic subclasses have been proposed, including hydrogen-rich, helium-rich, and spectroscopically featureless TDEs \citep[e.g.][]{TDE_spectra,Featureless}. This spectral diversity introduces significant uncertainties in the classification of new events \citep{Charalampopoulos_22}. 
	%AGN flares

Another major contributor is flaring in already active nuclei. AGN are powered by accretion onto a central SMBH \citep{Salpeter} and vary stochastically in the optical, and only occasionally producing flares of more than half a magnitude above the baseline and lasting months \citep[e.g.][]{AGN_flares_2, AGN_flares}. The mechanisms behind these are heterogeneous and largely unresolved, with proposed channels including sudden changes in the accretion flow, TDEs in active galaxies \citep[e.g.][]{Wyrzykowski2017, Frederick2021, Huang2023} or superluminous supernovae \citep{Drake_2011}. In some cases, these flares sometimes produce important changes in the spectral characteristics of the AGN itself, as is the case of CLAGNs \citep[e.g.][]{Ricci, Yang, Camus2026}.

	%Other NTs
	The boundaries between these classes are not sharp. Nuclear flares can show mixed photometric and spectral signatures, thus challenging the classification of these events \citep[e.g.][]{Pessi_CLAGN, asassn15_SN, assasn15_TDE, Wiseman2025}. 
    Recently, BFFs have emerged as a new class of nuclear transients, recognizable by prominent, broad He$\,$\textsc{ii}\,$\lambda4686$, N\,\textsc{iii}\,$\lambda4640$, and O\,\textsc{iii}\,$\lambda3133$ lines, produced by the Bowen mechanism and requiring a strong extreme-ultraviolet radiation field \citep{BFF}. BFFs show, however, smaller line velocities than the very broad ($>$ $10{,}000~\mathrm{km\,s^{-1}}$) Bowen lines sometimes found in TDEs. These objects are found in galaxies with pre-existing nuclear activity and decline slowly, over years \citep{Marzena}, and have been attributed to episodes of enhanced accretion rather than to stellar disruption. The mechanism powering these flares remains uncertain, and only a handful of examples have been reported, mostly in active galaxies \citep[e.g.][]{Marzena, 2025A&A...701A.224B}. 

    In this work, we approached the question of what flares in galactic nuclei empirically, selecting nuclear transients from the ZTF alert stream in real time by their location alone so that the full diversity of SMBH-related flares, from TDEs to the rarer ambiguous transients, can enter the sample alongside the supernovae that dominate it. We find that simple, early-time observables, principally colour and rise time, already carry much of the power to separate SMBH-related flares from this supernova background, with TDEs emerging as the clearest case. These results point toward selection strategies that will matter most in the LSST era, when alert volumes will far outstrip the capacity for spectroscopic follow-up and the early, photometric identification of SMBH activity becomes essential.

The paper is structured as follows. Section~\ref{sec:methods} describes the alert-stream filtering criteria, the real-time candidate selection and classification procedure. Section~\ref{sec:results} describes the resulting sample: its spectroscopic composition, the photometric properties that can separate TDEs from the supernova background, the recovery of known TDEs by the filter, and the performance of three machine-learning classifiers. Section~\ref{discussion} discusses the lessons for the design of nuclear-transient filters and their implications for the LSST era, and Section~\ref{sec:conclusions} summarises our conclusions. The three ambiguous events identified during the search are analysed separately in Paper~II (Concepci\'on et al., in prep.).

\section{Methods}\label{sec:methods}

	\subsection{Alert stream filtering criteria}
 We compiled a sample of nuclear transients reported to the TNS and selected in real time from the ZTF alert stream, with the aim of retrieving spectroscopic classification for as many candidates as possible. ZTF runs from the 1.2-m Samuel Oschin telescope at Palomar Observatory and has surveyed the northern sky since 2018 in the $g$,$r$ and $i$ optical filters reaching a limiting magnitude of 20.4 in $r$ in its 30-s exposures.\footnote{\url{https://www.ztf.caltech.edu/ztf-camera.html}} With a field of view of $47~\mathrm{deg}^2$ and hundreds of alerts issued per night,\footnote{\url{https://www.ztf.caltech.edu/new/ztf-hits-10000-supernova-discoveries.html}} it has become a primary source for the discovery and monitoring of transients and provides an excellent testbed for LSST. The large volume of transient alerts has motivated the development of community alert brokers, which ingest real-time alert streams from ZTF, enrich them with contextual information, and serve them to users for further analysis. We carried out our selection using the Lasair broker \footnote{https://lasair-ztf.lsst.ac.uk/} \citep{Lasair}, which provides programmatic access through an Application Programming Interface (API) and the ability to define and run custom filters on the live stream. Lasair also includes the contextual classifier Sherlock, which provides spatial cross-matching for transients and offsets from their associated host, as well as a predicted transient class out of seven possibilities that can be used for further filtering. We implemented our selection as such a filter, run as a Jupyter notebook through Lasair's API, applying the criteria described below.

This work was carried out within the framework of the ePESSTO+ survey (Public ESO Spectroscopic Survey of Transient Objects; \citealt{Pessto}), a large programme using the 3.6m New Technology Telescope (NTT) at the La Silla Observatory of the European Southern Observatory (ESO). Our project covered a total of ePESSTO+ 127 observing nights at this facility. This observing time was organized in 41 campaigns, each typically consisting in 2 or 3 consecutive nights with the same observers, which would be relieved by a new team for the next run.

We use the term "nuclear" to denote a transient spatially coincident with the centre of its host galaxy, where the central SMBH resides. Nuclearity is the only physical property on which our selection rests, since it is the natural proxy for a possible association with SMBH activity, and we deliberately imposed no cuts on colour, light-curve shape, or other contextual information, so that the full diversity of SMBH-related flares could enter the sample. In practice, we required candidates to lie within $0.5\arcsec\ $ of the cataloged centre of their host, given ZTF's typical centroiding accuracy of $\simeq 0.3\arcsec\ $ as a result of its pixel scale of $1\arcsec\ $ per pixel \citep{ZTF_Masci}.

We stress that nuclearity is a necessary but not sufficient condition for an SMBH origin: supernovae projected onto, or occurring within, the dense central regions of galaxies can also appear nuclear, while a small number of genuinely SMBH-related events may be offset from the photometric centre, for example those associated with recoiling or secondary black holes \citep{2024tvd}.
Beyond nuclearity, we applied a small set of practical criteria. We required candidates to be brighter than 19.5 mag in either the $g$ or $r$ band at the time of selection, set by the faintness limit at which the NTT yields a reliable classification. We note here that transients could be selected at any point in their lightcurves. To reduce contamination from Galactic sources, we excluded objects within 10\degr\ of the Galactic plane and rejected likely stellar sources, defined as those with a star–galaxy score (s-g score) above 0.5, as calculated by the Lasair broker. The s-g score ranges from near 0 for extended, galaxy-like sources to near 1 for point-like sources. Events with no identifiable host (labeled as "orphan" by Sherlock) were also discarded. We requested objects to have at least 3 detections, in order to avoid spurious candidates. To preferably select rising or near-peak events while they remained observable, we kept only those discovered within the previous 30 days to the run of the filter, taking the date of report to the TNS as the discovery date. Finally, although ZTF covers the whole northern sky, we retained only events below a declination of +35\degr, observable from La Silla Observatory.

    \subsection{Real time candidate selection}
	Our selection ran over a 16-month period, from 2024 June 6 to 2025 September 22. During the 41 ePESSTO+ observing campaigns in this window (totaling 127 nights), the filter was run daily; between campaigns, we ran it less frequently. In total, the filter was run 87 times in this period. This returns an average of one run every 5.5 days, with occasional gaps due to Lasair or ZTF downtime. For each run, the surviving candidates were inspected through a custom web interface that we developed and hosted at our institution. For every object, it displayed the ZTF lightcurve, the latest difference-image cutouts, and the available host-galaxy and catalog information, allowing each candidate to be assessed in its full photometric and contextual setting. A small team used this interface to vet obvious non-transient candidates \footnote{By non-transient candidates we refer to likely AGN or variable star candidates, which are sometimes flagged as new objects with their own IAU/ZTF names even if there is no obvious flare.}  or unobservable objects, record comments, and decide which objects to report to ePESSTO+ as classification targets, with priority given to bright events and to those still rising or near peak. Registering a candidate and communicating it to the observers  did not guaranty its observation. From the pool of reported targets, the ePESSTO+ observers selected which to observe according to visibility, weather, instrument availability, and the collaboration's priorities, which go beyond nuclear transients. As a result, only part of the candidates we reported were ultimately classified, and the final sample reflects these additional selection layers beyond our filter and vetting; we assess the resulting effects in Sect. \ref{discussion} . Maximizing the number of observed candidates from those that were returned by the filter was also a priority to minimize biases. In most cases, classifications are based on a single spectrum, although follow-up spectra were requested for a few cases.  Spectra obtained by ePESSTO+ were reduced using the PESSTO pipeline\footnote{https://github.com/svalenti/pessto} \citep{Pessto}. Classifications were reported to the TNS based primarily on spectral matching with  \citep{SNID} and SNID-SAGE \citep{Stoppa2026}, and occasionally complemented by the observed light-curve behaviour. Public classification reports are available on the TNS.  A further fraction of the sample was classified by other independent teams, for which we retrieved the public classifications reported to the TNS using the TNS API \citep{tns_api} and downloaded the spectra from the public repository for spectra Wiserep\footnote{https://www.wiserep.org/} \citep{wiserep}. Photometric analysis was done using data from ZTF, queried and retrieved directly from the Lasair broker via its API.

\section{Results}\label{sec:results}
\subsection{Statistics of the classified sample}
Our nuclear transient filter yielded 476 unique candidates over the selection period.  Individual runs averaged 28.75 objects, or roughly one new candidate per day given that each run only showed events discovered in the previous 30 days. Of these, 202 objects (42\%) were subsequently classified spectroscopically and reported by us or others to the TNS. This classified sample represents the primary focus of our analysis. This section is organized as follows. First, we describe the statistics of the classified sample in Sections \ref{res_1} and \ref{res_2}, by number of objects and by classifying group. In Section \ref{res_3} we show the results of an analysis of the properties of the lightcurves of the objects in order to study parameters which could improve filtering of nuclear transients in the future. A similar analysis on the part of the sample that was not classified is shown in Section \ref{non_class}, in order to study possible selection and observational biases. We include a more detailed description of the recovery of TDEs by our filter in Section \ref{res_4}, and we identify those that were missed and the reason why they were not found by the filter. In the same light, we also compare the performance of three available machine-learning based classifiers in Section \ref{sec:ml}. Finally, three candidates (AT2023zgo, AT2024eff and AT2024zkr) were identified and were deemed interesting for spectroscopic follow-up due to their ambiguous characteristics. Two were selected in test runs of the filter before the start of the project and are not included here, while AT2024zkr is included and we classify it as a BFF. These three events are analysed in detail in a companion paper (Concepción et al., in prep., Paper II).

\subsubsection{Composition of the classified sample and statistics}
\label{res_1}
\begin{figure*}[t]
  \centering
  \includegraphics[width=\textwidth,height=0.4\textheight,keepaspectratio]{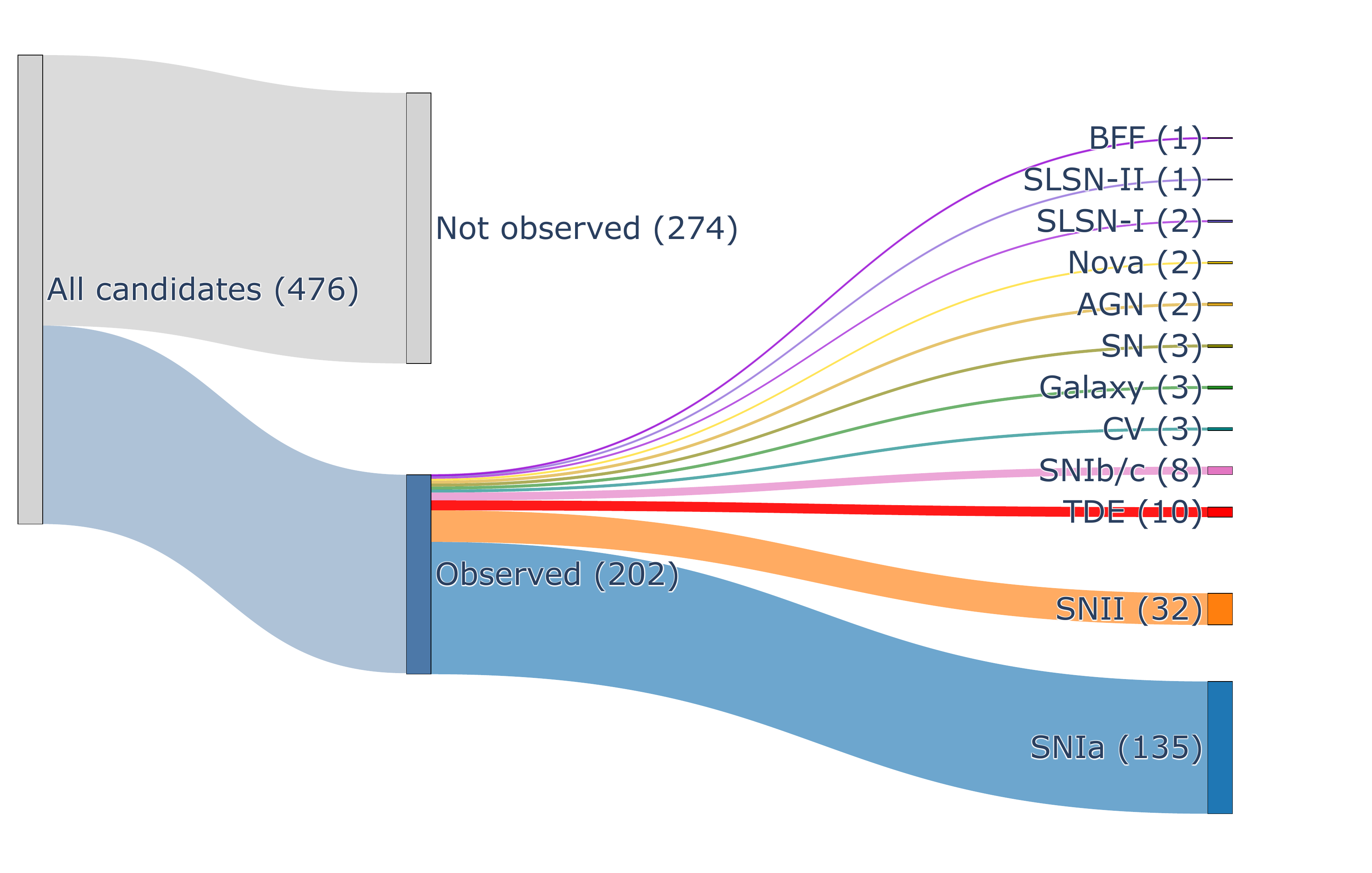}
  \caption{Transient census of the 476 filter-selected candidates, 202 (42\%) were spectroscopically classified. The classified sample is dominated by SNe Ia (67\%), followed by Type II SNe (16\%), Type Ib/c SNe (4\%), and TDEs (5\%). Objects classified as SN had low-quality spectra, which did not allow a reliable classification of the supernova type. The class Galaxy includes objects for which no distinctive transient features were identified against the host background, indicating the event was already too faint when it was observed. Only three Galactic objects (cataclismic variables, CV) entered the sample.}
  \label{fig:distribution}
\end{figure*}

Figure~\ref{fig:distribution} summarizes the breakdown of the 202 classified transients, based on the classification on the TNS. Type Ia supernovae dominate the sample, comprising 135 objects (67\%). Type II supernovae represent the second-largest group with 32 events (16\%), while Type Ib/c supernovae account for 8 objects (4\%). We found a total of 10 TDEs (5\%) during this period, along with 1 BFF (0.5\%). The remainder is shared among small numbers of superluminous supernovae, AGN, galaxies (where no transient could be identified above the spectrum of the host) and galactic objects. 

\subsubsection{Spectroscopic classification contributions}
\label{res_2}
\begin{figure}[t]
\centering
\includegraphics[width=\hsize]{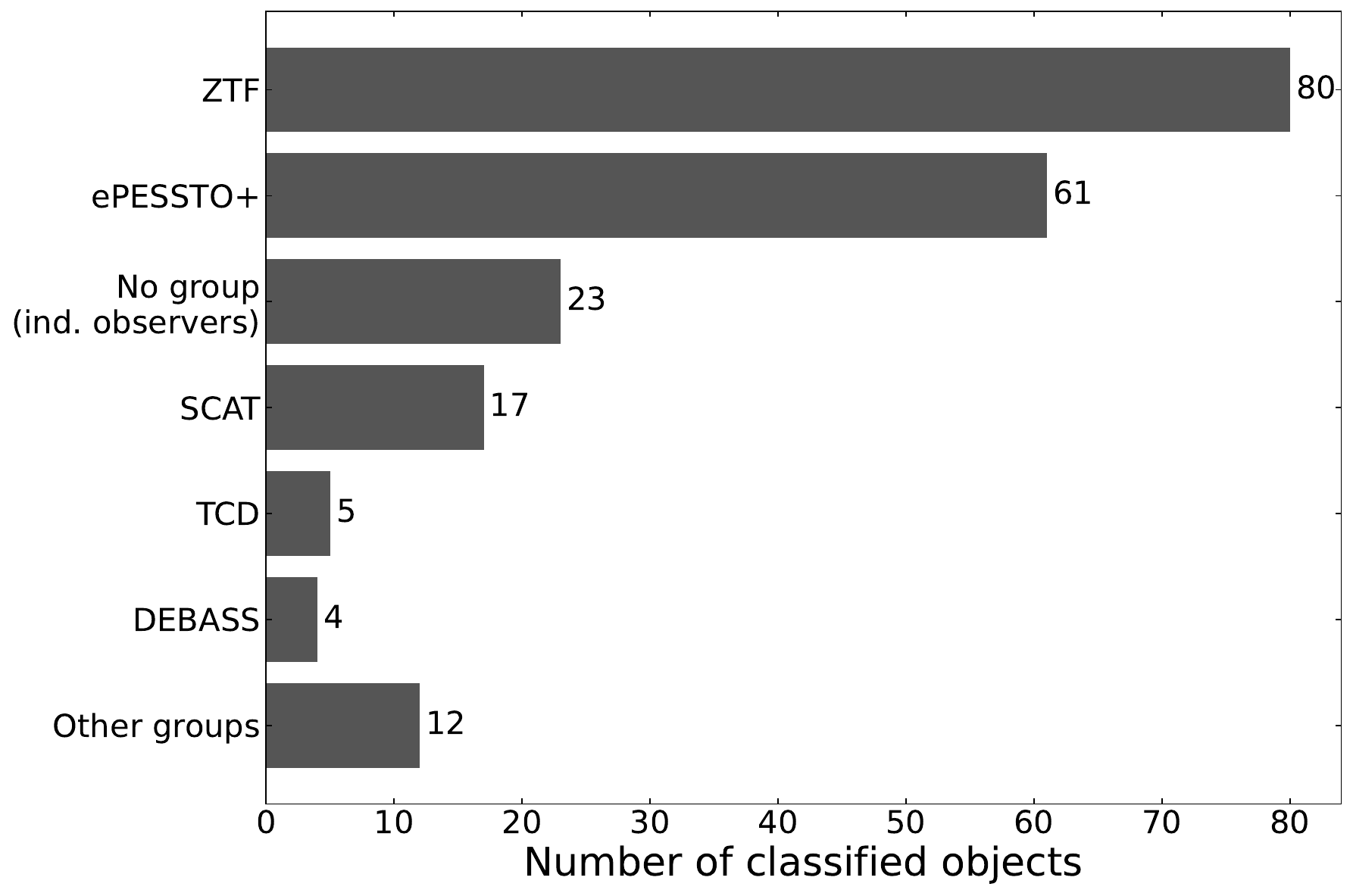}
    \caption{Classification contributions by group, as registered in the TNS. ePESSTO+ provided 30\% of the total spectroscopic classifications, with the remaining 70\% from other groups, led by ZTF. In "Other groups", we include FLEET (2), UCSC (2), YSE (2), RAPAS (1), SGLF (1), XOSS (1), ATLAS (1), StarDestroyers (1) and GOTO (1).}
    \label{fig:classifiers}
\end{figure}

Our aim was to obtain classifications through the ePESSTO+ group and a total of 61 events (30\%) were classified by this group, either on our direct request or by suggestion from other members of the collaboration. The remaining 70\% came from external groups that made their classifications public on the TNS,  with the ZTF group contributing the largest fraction at 80 classifications (38\% of external follow-up). A concrete breakdown of the contribution of different groups is shown in Fig.~\ref{fig:classifiers}.  

In order to asses the efficiency of our project, we gathered the numbers of all classified objects in the TNS (i.e. not focusing on nuclear transients, but the general transient population) per group in same time period. We found that ZTF is, by far, the most prolific classifier with 1338 total classifications, followed by ePESSTO+, with 581 and the Spectroscopic Classification of Astronomical Transients  \citep[SCAT;][]{SCAT} group with 439. In our project, the relative contribution of ePESSTO+ in nuclear transient classification is much more significant and comparable to that of ZTF, which highlights the efficiency of dedicated nuclear transient classification and follow-up programs within such collaborations.

\subsection{Photometric properties and transient classification}
\label{res_3}
We examined the photometric properties of the sample to assess the prospects for pre-selecting SMBH-related candidates, chiefly TDEs. The comparison is restricted to TDEs and supernovae: supernovae are the dominant contaminant and their lightcurves are morphologically comparable, whereas those of AGN and variable stars are not.

\subsubsection{Mean colour and colour evolution as a photometric discriminant}
\label{mean colour section}
We retrieved $g-$ and $r-$band ZTF difference photometry lightcurves from the Lasair broker for all classified transients in our sample. We calculated the mean $g-r$ colour for objects with at least 4 nights with both $g$ and $r$ detections, employing this conservative approach to avoid model-dependent interpolation and associated uncertainties in sparsely sampled lightcurves. Colours were corrected for Galactic extinction using the SFD dust map \citep{schlegel}, rescaled by 0.86 following the recalibration of \cite{Schlafly}, and the  extinction law with $R_V$ = 3.1 \citep{fitzpatrick}. 
 In total, 135 classified events from the sample had enough datapoints to perform this analysis. Mean $g-r$ colour emerges as a strongest photometric discriminant between TDEs and the dominant supernova classes. TDEs exhibit predominantly blue colours, with a mean of  $g-r=-0.25\pm0.05$ mag (10 events). This contrasts with all supernova types: SNe Ia are marginally positive at 0.08 ± 0.02 mag (92 events), while Type II and Ib/c SNe are substantially redder at 0.35 ± 0.05 (24 events) and 0.49 ± 0.19 (6 events) mag, respectively. Only two SLSNe had enough points for the analysis, and they return slightly blue colours, -0.06 ± 0.11. 

\begin{figure}[t]
\centering
\includegraphics[width=\hsize]{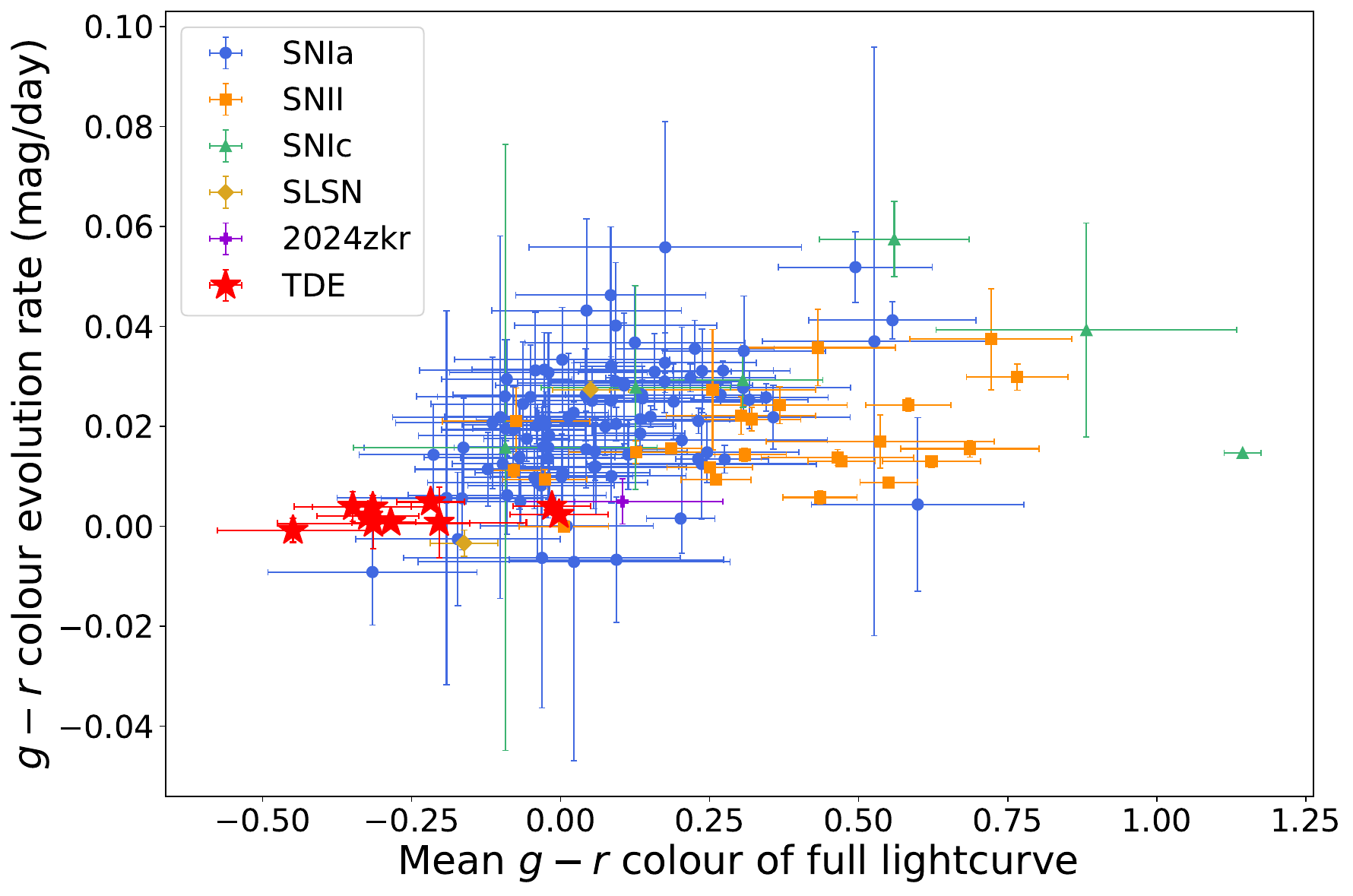}
    \caption{Linear evolution of the $g-r$ colour, in magnitudes per day, versus the mean $g-r$ of the full lightcurve. TDEs (red) evolve toward redder colours more slowly than supernovae, providing a potential temporal discriminant. A linear fit is applied to each object using all available photometric same night $g$ and $r$ points per object, in order to obtain the slope which reflects the colour evolution trend.}
    \label{colour evolution}
\end{figure}

The rate of colour evolution provides a complementary diagnostic. For the same full, well-sampled lightcurves (again, events with $\geq 4$ same-night $g$ and $r$ detections), we fit linear trends to $g-r$ as a function of time and quantify the reddening rate. TDEs redden remarkably slowly: 0.0022 ± 0.0006 mag/day. This is approximately one order of magnitude slower than all supernova types, which range from $0.0120$ to $0.0339$ mag/day. Stripped-envelope supernovae show the fastest evolution. The slow colour evolution of TDEs relative to SNe has been noted previously . \citep[e.g.][]{VV2011, Seventeen}. Here, we quantify this effect in a sample selected in real time using uniformly measured ZTF $g-r$ colours and demonstrate its potential as a simple temporal discriminant for real-time selection from the alert stream . Figure~\ref{colour evolution} shows the colour evolution of these objects compared to their mean colours, showing a significant separation of TDEs from the main bulk of SNIa.

As early detection of candidates is key, we repeat the analysis focusing only on pre-peak lightcurves, calculating the mean $g-r$ colour of objects before the maximum brightness. We do this for objects with now 3 same-night $g$ and $r$ detections before peak, since the number of detections is now more limited. We define the time of peak as as the epoch of maximum flux, carefully excluding spurious detections where the point is more than 0.3 magnitudes above the previous and following detections. All transient types are bluer (smaller or more negative $g-r$) in these early stages, but the relative ordering is preserved and the TDE–SN separation remains significant. Table \ref{tab:colours} shows the average full-lightcurve mean $g-r$ and pre-peak $g-r$ for each class. A figure similar to Fig.~\ref{colour evolution}, but focused on pre-peak detections is included in the Appendix (see Fig.~\ref{colour evolution pre})

\begin{table*}[t]

\centering
\caption{Photometric characteristics of TDEs and supernovae in the classified sample.}
\renewcommand{\arraystretch}{1.3}
\label{tab:colours}
\begin{tabular}{c c c c c c p{4cm}}
\hline\hline
Class & Mean peak mag & Mean $g-r$ (pre-peak) & Mean $g-r$ & $\Delta(g-r)$ (mag/day) & Median rise time (days)\\
\hline
TDE  & 18.05 $\pm$ 0.23 &-0.26 ± 0.04 & -0.25 ± 0.05  & 0.0022 ± 0.0006 & 34.0 ± 6.1 \\
SNIa & 18.26 ± 0.05 & 0.04 ± 0.02  & 0.08 ± 0.02 & 0.0206 ± 0.0013 & 16.5 ± 1.4\\
SNII & 17.90 $\pm$ 0.15 & 0.30 ± 0.06  & 0.35 ± 0.05  & 0.0174 ± 0.0019 & 15.9 ± 2.6\\
SNIb/c & 18.14 ± 0.24 & 0.35 ± 0.17  & 0.49 ± 0.19 & 0.0339 ± 0.0070 & 20.9 ± 3.6\\
SLSN & 17.91 ± 0.35 &-0.12 ± 0.06 & -0.06 $\pm$ 0.11 & 0.0120 ± 0.0153 & 15.4 ± 9.4\\ 

\hline
\end{tabular}
\end{table*}

%\FloatBarrier

\subsubsection{Rise time proxies as a morphological discriminant}
We used the time between first light and peak brightness as a proxy for the rise time of the transients in the classified sample. As in \citet{Galbany2025}, we used forced photometry to identify flux below the alert-detection threshold and defined $t = 0$ from the last constraining non-detection. Unlike \citet{Galbany2025}, who adopted the midpoint between the last non-detection and the first detection or a fit to the early rise, we set $t = 0$ at the last non-detection itself, in order to avoid the uncertainties arising from fitting the same rise model to different classes of transients.
We also lowered the detection threshold from $S/N \geq 3$ to $S/N > 2$ to account for the heterogeneous sampling of the light curves. Specifically, we identified the first forced-photometry point with positive flux and $S/N > 2$ within the five days preceding the first alert detection. We assumed that this point was already on the rise, and set $t = 0$ at the preceding forced-photometry point or constraining upper limit. We then measured the rise time as the interval between $t = 0$ and the peak, which was chosen as described in \ref{mean colour section}. Because $t = 0$ is set at the last non-detection, these values are upper limits on the true rise time within the cadence gap. We computed rise times only for events where both $t = 0$ and the peak were well constrained, which was the case for 174 events (86\% of the classified sample). We applied no time-dilation or stretch corrections, since we intend these parameters to be usable in real-time, redshift-agnostic searches.

We find that the rise time distributions provide an additional morphological discriminator between TDEs and supernovae. TDEs show the longest rise times in the sample, with a median value of 34.0 ± 6.1 days (10 objects). In comparison, SNe Ia exhibit significantly shorter rise times, with a median of 16.5 ± 1.4 days (128 objects), while Type II and Type Ib/c SNe have a median of 15.9 ± 2.6 days (26 objects) and 20.9 ± 3.6 days (8 objects), respectively. The SLSN sample only counted with 2 objects which showed an average rise time of 15.43 ± 9.4 days.

These values are broadly consistent with literature rise times for SNe Ia \citep[e.g.][]{rise_Ia_1, rise_Ia_2} and stripped-envelope supernovae \citep[e.g.][]{str_env_1, str_env_2}, with the small deviations likely due to the lack of corrections mentioned above and the inherent sampling and magnitude limitations of the survey. We find that SNe II show longer rise times than the typically reported timescale of $\sim$7.5 days \citep{Rise_SNII}. We attribute this to the fact that most classifications on the TNS are based on a single spectrum and not analyzed in detail, thus grouping the longer-rising hydrogen-rich supernovae under the same category. In our sample, we find only one of each IIn, IIb and IIP types  being explicitly classified as such, but likely more are included. Approximately one third of the objects show rise times above 17 days, which has been suggested as a possible threshold between type II and IIb \citep{2b}. Regarding SLSNe, these objects show a wide range of rise times, sometimes extending to much longer timescales, but faster events like those seen here are also common \citep{SLSN_rise_times}.

\begin{figure}[t]
   \centering
   \includegraphics[width=\hsize]{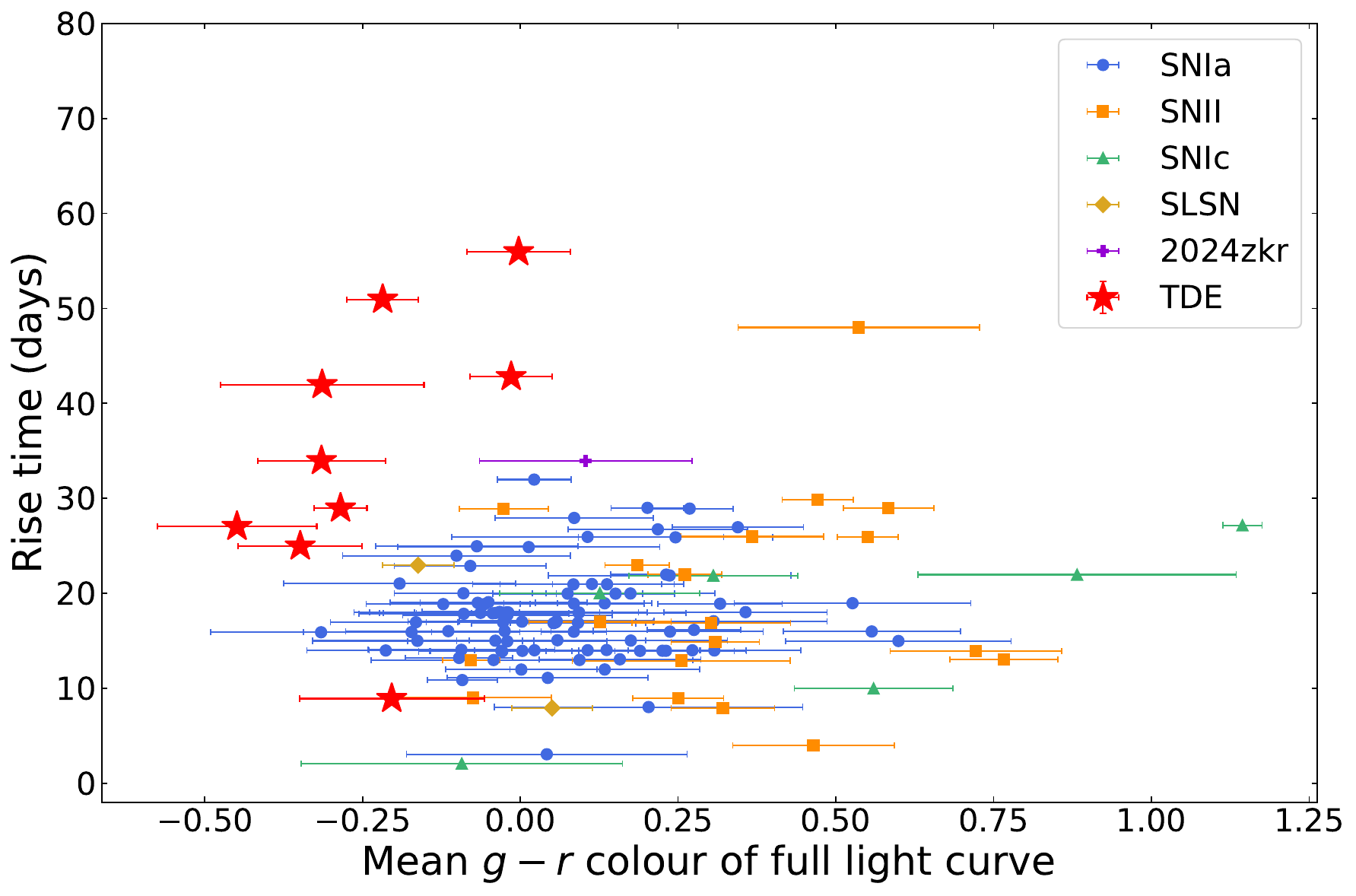}
      \caption{Rise time against mean $g-r$ colour. TDEs segregate to the upper-left (long rise times, blue colours), distinct from supernovae which cluster in the lower-right. Only TDEs and SNe are included; AGN and stellar variables lack well-defined rise times.}
         \label{rise_times}
   \end{figure}

Overall, TDEs show longer characteristic timescales than the dominant supernova populations, although there is a natural overlap due to the intrinsic diversity of transient lightcurves. Figure \ref{rise_times} displays a comparison of mean $g-r$ colour against rise times. TDEs occupy the upper-left region (longer rise times, blue colours), well separated from the cluster of supernovae.

\subsubsection{Peak apparent magnitudes, redshifts and nuclear offsets.}
\begin{figure}[t]
   \centering
   \includegraphics[width=\hsize]{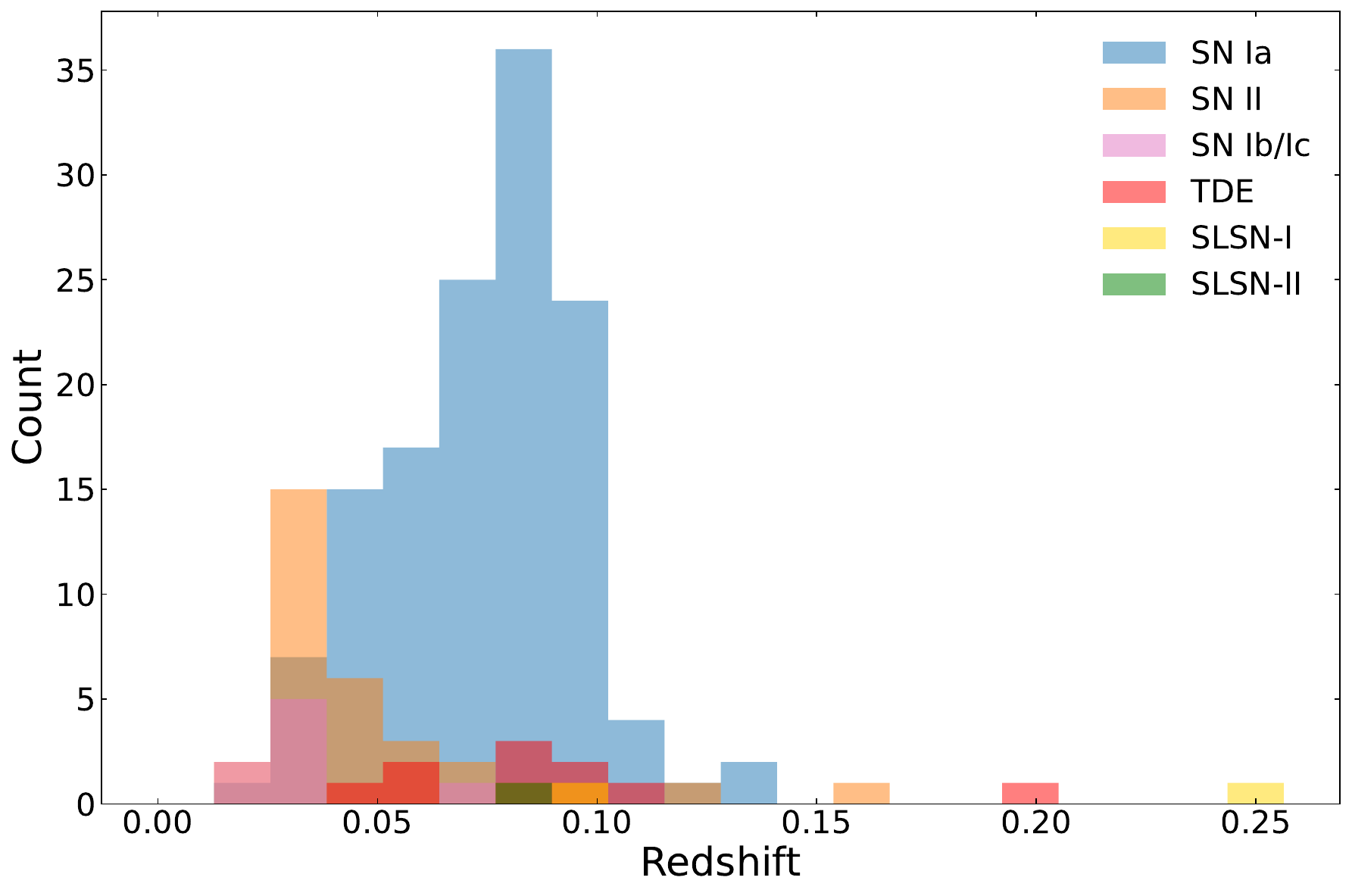}
      \caption{Redshift distribution of classified transients. Galactic objects are excluded. Most events lie at $z < 0.1$; Type Ia SNe peak near $z = 0.07$.}
         \label{redshift}
   \end{figure}

 We recovered the redshift, nuclear offset and apparent peak magnitude, in any band, for the objects in our sample. This serves two purposes. First, it allows us to compare the classified objects with those that remained unclassified (see Section~\ref{non_class}) and assess how brightness affected their likelihood of being registered and observed. Second, since LSST will probe a substantially larger volume than ZTF, understanding the redshift and brightness distributions of transients in current surveys provides a necessary baseline for interpreting those expected from LSST. We find a mean peak apparent magnitude across all events of 18.18 $\pm$ 0.06. Among true transients, SNe II show the brightest values at 17.90 $\pm$ 0.15, closely followed by  SLSNe with 17.91 ± 0.35, TDEs at 18.05 $\pm$ 0.23, and SN Ib/c at 18.14 ± 0.24. SNe Ia are the faintest group, showing a mean peak apparent magnitude of 18.26 ± 0.05. As expected given ZTF's magnitude depth and survey design, our transients lie at low redshift, with $z < 0.1$ for most of the sample. Only two objects exceed $z = 0.2$: one TDE and one superluminous supernova. Type II and Ib/c supernovae cluster at lower redshifts than SNe Ia, which peak near $z \approx 0.07$ (Fig.~\ref{redshift}), consistent with their differences in typical absolute magnitudes. 

Regarding their nuclear offsets, all 10 TDEs in our sample are localized within 0.35\arcsec\ of their host galaxy nuclei. This tight spatial concentration suggests our current nuclear selection criterion (0.5\arcsec) could be tightened to 0.35\arcsec\ , potentially reducing off-nuclear contamination. However, it was noticed during the study that this separation could vary slightly for the same transient with new detections, likely due to the observing conditions. Additionally, cases like off-center TDEs due to a secondary SMBH in the galaxy would be excluded with strict cuts. In contrast, supernovae show no systematic spatial-offset trends across subtypes.

\subsubsection{The non-classified sample}
\label{non_class}
In order to study possible biases in the classified sample, we repeated a similar photometric analysis on the part of the sample that was not spectroscopically classified. We obtain a mean $g-r$ colour of 0.13 $\pm$ 0.02 and a mean reddening slope of 0.013 $\pm$ 0.002 from a subset of 139 objects with enough same-night data points. Limiting the analysis only to pre-peak detections, the mean colour of the 91 events with sufficient data before peak is now 0.06 ± 0.02. These values are all strikingly similar to those of SN Ia (see Table~\ref{tab:colours}), which indicates the non-classified sample is likely also dominated by these objects. Figure \ref{fig:not_class_analysis} shows similar plots to those in figures \ref{colour evolution} and \ref{rise_times} for the part of the sample that was not observed. In the same way, we also calculate the rise times for 253 objects, and find a median value of $18.4 \pm 6.0$ days, which is again similar to the values we obtained for supernovae.

Three relatively distinct groups can be identified according to their rise times, as seen in Fig. \ref{fig:not_class_analysis}. Those with more than 100 days are all flagged as AGN by Lasair, and don't show signs of distinct flares above their natural variability. A smaller group appears between 40 and 100 days, where some genuine, longer rising flares are mixed with other AGN candidates. The most numerous group is clumped at shorter rise times, showing a distribution very similar to the one of the classified sample and again indicating it is likely dominated by supernovae.

This non-classified sample is significantly fainter than the classified events by around 0.5 mag, showing a mean peak magnitude (in any band) of 18.68 $\pm$ 0.05. This confirms the observational bias towards brighter events, which were more likely to be registered and observed.

\subsection{TDE recovery and completeness}
\label{res_4}
To quantify the completeness and selection bias of our nuclear transient filter, we cross-matched our sample against all TDEs formally reported to the TNS within our target declination range and observation period. A total of 25 TDEs met these criteria; our filter included 10, yielding a 40\% recovery rate among all classified TDEs (as reported in the TNS) in the relevant time period. The recovered TDEs are listed in Table~\ref{table_colours} with their photometric properties, and their detailed ZTF lightcurves and classification spectra are shown in the Appendix. Among these, we find one event classified as TDE-H-He, two TDE-He and eight classified only as TDE. We note that the classification of TDE2025cyh, which was classified as TDE with no given subclass, is ambiguous, as it also shows narrow AGN-like line ratios along with H$\beta$ and \ion{He}{ii}/\ion{N}{iii}. It was classified as a possible TDE candidate as it did not show the typical AGN colours and variability in infrared data, but classifiers also note the possibility of it being a Bowen Fluorescence Flare \citep{2025TNSCR.992....1H}. SNID-SAGE agrees with the TDE classification, providing a best match with TDE~2019ahk \citep{ahk}. No public follow-up spectrum is available for further analysis.

For the 15 missed TDEs, we investigate the reasons why the filter did not include them. These are enumerated below.

\begin{itemize}
\item \textbf{Magnitude limit} (2025spf, 2025hee, 2024agqx, 2025fak): Four TDEs never reached our 19.5 mag detection threshold, and as such never passed the filter. We do not interpret this as a failure of the filter, but rather as a consequence of spectroscopic observational limitations.

\item \textbf{Star-galaxy score null values} (2024wsd, 2024ule, 2024admt): The Lasair broker returned null star-galaxy scores for these events, meaning no numerical score was available at all. Upon finding these missed events, we modified the filter to accept null scores.

\item \textbf{Broker host-galaxy mismatch} (2025ug, 2024ymu): The automated cross-matching in Lasair returned offsets of $24''$ and $19.7''$, respectively, exceeding our 0.5 arcsec threshold. Visual inspection of archival imaging, however, clearly positioned both events at or very near the galaxy nucleus. These represent failures of the automated host-galaxy association algorithm, not true spatial offsets.

\item \textbf{Object not available in Lasair} (2025iuh): this transient did not appear on the Lasair broker, and we were not able to search for it neither by its IAU name nor by its ZTF name. 

\item \textbf{Time-window constraint} (2025cyj): This event exceeded our 30-day threshold between first detection and brightening above the 19.5 mag trigger level. The extended pre-trigger baseline placed it outside our temporal selection window.

\item \textbf{Off-nuclear TDE} (2024tvd): This event showed a separation from the center of its host of 0.9" \citep{2024tvd}, thus falling outside of our sample. It was identified as the first optically selected off-nuclear TDE, likely due to a stellar disruption around a secondary SMBH in its host galaxy.
 
\item \textbf{Gap in filter execution} (2025axp): The ePESSTO+ observing runs scheduled between late February–April 2025 were removed from the schedule, so the filter was not run for a significant time period and this object was missed.
\end{itemize}

Finally, two events (2025kkj and 2025vjw) were not recovered, and we could not identify the specific criterion responsible for their exclusion. We speculate that this was due to one of the filtering parameters, such as the host offset or star–galaxy score, having changed between the time the transient was active and our subsequent analysis. It is also possible that the TNS crossmatch service was not running correctly during this period.

\begin{table*}
\centering
\caption{TDEs reported to the TNS that were recovered by our filter in the
relevant time period, with the mean $g-r$ colour over the full light curve and
before peak, the rise time, and the predictions of the NEEDLE, FLEET and ALeRCE
classifiers. A dash indicates that no prediction was available; ``n.e.p.''
indicates too few pre-peak detections for a colour measurement.}
\label{table_colours}
\centering
\renewcommand{\arraystretch}{1.3}

\begin{tabular}{c c c c c c c c p{3cm}}
\hline\hline
ZTF name & TNS name & Mean $g-r$ & Mean $g-r$ (pre-peak) & Rise time (days) & NEEDLE & FLEET & ALERCE\\
\hline\hline
ZTF25aaymnha & 2025qhf & -0.28 ± 0.04 & -0.23 ± 0.02  & 29 & TDE & - & TDE \\
ZTF25aajfjhm & 2025eve & -0.31 ± 0.16 & Not enough points & 42 & TDE & TDE & TDE\\
ZTF25aagevje & 2025chm & -0.32 ± 0.10 & -0.31 ± 0.05& 34 & TDE & TDE & TDE \\
ZTF25aafywpr & 2025ccg & -0.20 ± 0.14 & -0.17 ± 0.11  & 9 & SLSN-I & - & SNIa\\
ZTF25aafwfzz & 2025cyh & -0.35 ± 0.09 & -0.30 ± 0.06 & 25 &- & -  &TDE\\
ZTF25aafofcs & 2025bri & -0.01 ± 0.07 & -0.22 ± 0.04 & 43 & TDE & - & TDE\\
ZTF24abyhjvc & 2024aepd & -0.45 ± 0.12 & Not enough points & 27 & unclear & TDE & TDE\\
ZTF24abmybnp & 2024yqo	& 0.00 ± 0.08 & -0.07 ± 0.05 & 56 & TDE & TDE & SNIIn\\
ZTF24abfuwbm & 2024vip	& -0.32 ± 0.08 & -0.38 ± 0.10 & 42 & - & - & TDE\\
ZTF18abxnvoz & 2024pvu & -0.22 ± 0.06 & -0.22 ± 0.02& 51 & - & TDE & TDE\\

\\
\hline
\end{tabular}
\end{table*}

\subsection{Machine-learning classifier performance in TDE selection}
\label{sec:ml}

Machine-learning classifiers are increasingly used to identify transients from photometric and contextual data, and will be essential at LSST data volumes \citep[see e.g.][]{Mallorn}.  A common limitation is the scarcity of confirmed TDEs for training relative to the much larger supernova and AGN populations, which increases contamination and degrades performance at early times, when reliable classification is most valuable for triggering follow-up. Several classifiers are already available, each with their own unique scope and approach. Here we recover the scores of three TDE classifiers and assess their performance against our sample.

The Finding Luminous and Exotic Extragalactic Transients (FLEET) algorithm \citep{FLEET1, FLEET} is a random forest built to select a single class in a redshift-agnostic manner and optimized for sample purity. Originally aimed at SLSNe, it later included an expansion targeting TDEs. It uses parametric lightcurve fit parameters, together with host-galaxy contextual information such as the nuclear offset. Its best performance occurs around 40 days, performing better than in the early stages of the lightcurve and finding no significant improvement afterwards.

Similarly, the NEural Engine for Discovering Luminous Events (NEEDLE) \citep{Xinyue, Xinyue2} aims at exploiting the host-galaxy properties for classification, as TDEs typically prefer green valley post-starburst galaxies with masses below a few $10^{10} \, M_{\odot}$ \citep{TDE_hosts}. A convolutional branch processes the science and reference image cutouts, while a dense branch processes host-galaxy photometry, the nuclear offset and a few simple light-curve features from the alert packet. It classifies events as SLSN, TDE or SN, with an additional "unclear" category. A prediction can be issued even with one detection, although for the first two classes the probability typically increases as the lightcurve approaches its peak.

Finally, the ALeRCE broker \citep{Alerce_broker} offers a Balanced Hierarchical Random Forest (BHRF) classifier, which processes all public ZTF alerts and recently included a TDE subclass for its classifications \citep{Alerce1, Alerce2}. A first level divides the events into transients, stochastic and periodic objects, while a second random forest is applied to each of the previous classes and classifies into subclasses, with TDEs under the transient category. It includes 152 general features plus another 24 specifically aimed at distinguishing TDEs. Among these are the decay value obtained from fitting a simple model to the lightcurve and other features derived from the parametric fits done by FLEET.

We examined the performance of NEEDLE, FLEET and ALERCE against our spectroscopically confirmed sample; predictions for the TDEs are included in Table \ref{table_colours}. We stress here that these scores were recovered after all the objects had been classified and faded away, so these predictions are done on the whole lightcurve.

NEEDLE is available on the Lasair broker as an annotator \footnote{https://lasair-ztf.lsst.ac.uk/annotators/NEEDLE/}, allowing users to crossmatch their filters with NEEDLE's classifications. We recovered the scores from the latest available version of the classifier on Lasair. This annotator applies pre-selection filters before its pipeline, so not all classified events received a prediction: it returned scores for 130 of our 202 classified transients. Among the 10 confirmed TDEs, NEEDLE correctly identified 5, assigned no score to 3, labelled 1 as "unclear" and misclassified 1 as a superluminous supernova. 

FLEET\footnote{https://fleet-pipe.readthedocs.io/en/master/} scores were available for only 100 transients and were recovered through its Python package, which uses the 3.0 version of the classifier. In some cases, the lightcurves did not contain enough points for fitting, while in other cases the catalog matching failed. Of those that received a prediction, seven had TDE as the most probable class, five of them correctly and with confidence above 80\%. The remaining two, 2024zkr and 2024zxw, had TDE as the best class but with lower probabilities (29\% and 61\%); 2024zkr we classify as a BFF and will be examined in detail in Paper II, while for 2024zxw a spectrum obtained by ePESSTO+ roughly one month after peak, at magnitude 20, and shows no features beyond host-galaxy emission. 

For ALERCE's BHRF classifier we recovered the classifications via its API\footnote{https://alerce.readthedocs.io/en/latest/apis.html} querying for "lc\_classifier\_BHRF\_forced\_phot" or "lc\_classifier\_BHRF\_forced\_phot(beta)". 
We retrieve the highest probability prediction from the latest available.
In most cases, this was version 2.1.0 or 2.0.0, with 1.1.5 and 1.1.4 in a few cases. In total, it returned a classification for 171 candidates, with 42 events being predicted as TDEs. Among these were 8 of the 10 TDEs in our sample, with the remaining two being predicted as a type IIn supernova and a SNIa (both by the 1.1.5 version as the latest available). 

Two points follow from this analysis. First, where a score was produced it was usually correct, with very few misclassifications for all three classifiers. The dominant failure mode was coverage rather than accuracy. Both NEEDLE and FLEET depend on external host-galaxy photometry, which was unavailable or unmatched for 35\% and 50\% of the sample respectively. This is particularly critical for nuclear transients, where the host is precisely the object whose photometry is hardest to measure reliably at the position of the flare. FLEET has a very high purity, while ALERCE has the best completeness, although it significantly overestimates the number of TDEs in the sample.

Second, this bottleneck will become more acute with LSST, which will detect transients well beyond the depth of existing host catalogs. Host-agnostic classifiers operating on alert photometry alone are therefore likely necessary in the early survey stages \citep{Anilkumar2026}. The trade-off is that discarding host information also discards the context that separates a nuclear flare from ordinary AGN variability, leaving light-curve morphology to carry the classification alone. This biases selection towards events that resemble the training set and against atypical flares. Rapid spectroscopic classification therefore remains necessary, making consistent and efficient filtering a critical feature of nuclear transient searches.

\section{Discussion}
\label{discussion}
\subsection{Sample composition and comparison with a previous experiment}
Over a 16-month period, our filter on the Lasair broker recovered 476 unique transient candidates from the ZTF alert stream. Out of these, 202 (42\%) were classified spectroscopically, returning a sample dominated by Type Ia supernovae (67\%) and including 10 confirmed TDEs (5\%). The unclassified events, which account for 58\% of the total, were not observed for several reasons, primarily observability and brightness constraints. In some cases, visual inspection of the lightcurves was sufficient to identify likely variable stars or standard AGN variability within historical limits, rather than flares of interest, and these were thus not reported. In the LSST era, this will be much more complicated given the larger number of events, which highlights the importance of efficient filtering and early prediction to separate potential SMBH-related candidates for classification and follow-up. 

Among true extragalactic transients, the main discriminant between classified and unclassified transients is apparent magnitude and lightcurve phase rather than intrinsic type. As 70\% of transients in our sample were classified by other groups, this introduces another possible bias in the classified sample, for example, if these external groups were targeting blue objects. An analysis of the photometric characteristics of the non-classified candidates shows that these were on average fainter, but the mean values of colours and rise times seem to point to a population very similar to the one of the classified sample. We therefore expect our classified sample to be statistically representative of the true population of nuclear transients within the magnitude limit of our census. 
 
While our selection criteria were mostly practical and observational, we do note a physical bias against slow-rising transients as a consequence of our 30-day detection window. This criterion was set to preferentially select young, rising or near-peak transients in order to avoid old, unobservable candidates. This bias is particularly relevant when combined with magnitude constraints. Additionally, the ePESSTO+ collaboration, which contributed 30\% of classifications, generally excludes classifications in AGN hosts. Moreover, flares in AGN that are insufficiently distinct from the baseline may not be flagged as new transients. This introduces a bias against SMBH-related events in galaxies with pre-existing nuclear activity. Algorithms that are able to reject regular AGN variability while still selecting the significant flares is optimal, especially in the LSST era where AGN are expected to be a major contaminant in the search for TDEs \citep[e.g.][]{Magill26}. We stress here the need to include known AGN hosts in nuclear transient searches.

A similar program focused on selecting nuclear transient candidates in real time
was carried out by \cite{dgany2023needle}. While their aims were
comparable, our initial filter differs from their approach in several important aspects, affecting the composition of final sample. Firstly, they explicitly aimed to remove AGN candidates, in particular by rejecting candidates with detections older than 100
days ("objects.jdmin() $<$ 100", in Lasair’s schema) and removing
objects with names starting with "ZTF19". In our
study, we used the TNS-reported discovery date in order to select newly rising or near-peak transients, which maximized the
probability of observing and classifying an object, and also to
avoid transient mismatches caused by earlier detections at the
same position, which we noticed was common when using objects.jdmin(). This temporal criterion also proves more reliable
than filtering by object name, as many genuinely new transients
retain ZTF designations from earlier years (e.g. beginning with
ZTF19 despite being detected in 2024). We did notice, however,
a few cases where Lasair did not show the TNS name. Together,
these modifications reduced the number of new daily candidates
to approximately one, making a visual inspection step still
feasible. Our recovered 5\% TDE fraction is higher than the one
they recorded: 2.2\% in their sample of nuclear events within
0.5\arcsec\ and 4.1\% when restricted to blue objects. The amount of
missed TDEs is also lower, and we show that most of them would
have been recovered with minor modifications in the filter, instead of being a consequence of visual inspection.

\subsection{Filtering insights: TDE selection and completeness}
\label{sec:filtering_insights}

Several lessons emerged from running the filter in real time that can be
applied to the selection of SMBH-related transients from the general
population of nuclear events.

\begin{itemize}

\item \textbf{A discovery window of 40--50 days is sufficient.} This allows most TDEs to rise while adding few supernova contaminants, since these typically rise faster. Searches that allow for much longer time windows will be dominated by AGN \citep[e.g.][]{Lawrence_2016}.

\item \textbf{The TNS discovery date is an effective age cut.} Filtering on
the date reported to the TNS restricts the sample to genuinely new events and
suppresses spurious candidates, but it assumes prompt communication between
survey teams, the TNS and the brokers. 

\item \textbf{A magnitude limit is necessary but costly.} A brightness cut is unavoidable if follow-up is to be carried out with a specific facility, yet it also removes potentially interesting events. This can be better handled through parallel filters with different brightness criteria, rather than as a single hard cut.

\item \textbf{Machine-learning classifiers depend on what reaches them.}
NEEDLE and FLEET were largely accurate for the TDEs they scored, but a
substantial fraction of our sample never entered their pipelines and received
no score at all (Sect.~\ref{sec:ml}), while ALeRCE was more complete but overestimated the number of TDEs. The pre-filtering step is therefore as
important as the classifier itself, which is a further argument for
consistent and well-characterized filters upstream. Approaches that do not
require host-galaxy inputs altogether \citep[e.g.][]{Anilkumar2026} are attractive for
the LSST era, but their reported performance is derived from complete
archival lightcurves; applied to a live alert stream, where only the
observed portion of the lightcurve is available, performance will differ. Thus, early lightcurve classifiers are also essential \citep[e.g.][]{Early_TDEs}.

\item \textbf{Nuclearity depends on host association by the broker.} The
reported nuclear offsets are generally consistent, but incorrect host
matching excluded several TDEs from our sample. Since this association is
performed by the brokers, we recommend cross-checking between brokers or
against independent catalogs. A threshold of $0.5\arcsec$ proves reliable
in standard cases, but it excludes off-nuclear TDEs, and astrometric
precision is instrument dependent, so the same threshold may not transfer to
other surveys.

\item \textbf{Not rejecting active galaxies pays off.} The filter was not optimized to find flares in active nuclei, but neither did it reject them, and three events with non-standard characteristics were recovered and followed up spectroscopically (to be presented in  Paper II). Live filtering of this kind is therefore useful not only for TDE science but for SMBH-related transients more broadly. 

\item \textbf{Simple light-curve information can be added in the filtering process.} Our filter used no colour or light-curve criteria. As shown above, basic parameters such as
the mean $g-r$ colour, even before peak, and the rise time can be
incorporated as additional criteria to reduce supernova contamination already in the filtering stage.

\end{itemize}

Our filter recovers 10 of the 25 spectroscopically classified TDEs within the
relevant time window and available for classification with the NTT, given its declination and magnitude limits . Examination of the missed events
shows that the adjustments described above, namely removing the magnitude
limit, extending the first-detection window and admitting events with a null
s-g score, would raise the recovery to 19, while at least two
further losses are attributable to incorrect host association by the broker. 
Since the typical rise time for supernovae is already within the 30-day window used here, extending this period would not significantly increase the amount of supernovae, but only allow longer-rising events to enter the sample. At the same time, the s-g score is agnostic towards the event class, so admitting events with null score should also not alter the class fractions. A larger effect is expected when removing the magnitude limit, which would increase the number of events. Fainter or higher redshift candidates would have poorer host associations, potentially increasing the contamination from non-nuclear events. Additionally, given the higher rate of SNIa, the larger volume probed at lower magnitudes would likely increase the fraction of these events. This will have a larger effect in LSST, where the magnitude range is much wider. Combined with the light-curve properties discussed previously, this selection strategy offers a practical route to isolating TDEs in real time.

\section{Conclusions}
\label{sec:conclusions}

We have presented a real-time, nuclear-spatial filter applied to the ZTF
alert stream. The filter applied only spatial and practical constraints, with no photometric cuts and no assumptions about the host, and in particular did not reject active galaxies. It yields a TDE fraction of 5\%, higher than in previous nuclear transient searches. Modest adjustments to the filter show that a high recovery fraction of TDEs within the total known sample of these events is achievable. We further show that simple light-curve parameters measurable from the alert stream can be used to separate TDEs from their dominant contaminants even before peak, and can therefore be added as filtering criteria rather than applied only in hindsight.

This experience can be applied to Rubin Observatory's LSST, which is now delivering a substantially larger alert stream at fainter magnitudes. 
\citet{Bricman} predict 12-20 detectable TDEs per night, of order a
hundredfold increase over the current sample. Two consequences follow. First, spectroscopic confirmation of all of them will not be feasible, so any large Rubin TDE sample will necessarily be selected photometrically, and selection bias becomes as important as contamination: a photometrically selected sample reproduces the properties of its training set, and the events least likely to be recovered are precisely those that do not resemble it. Efficient filtering and well-characterized photometric classification therefore become essential, as does an honest account of what any given selection excludes.

Second, the Rubin population will be considerably fainter than the one currently known. The known optical TDEs peak at $r\approx18-19$, whereas the peak of the Rubin TDE distribution is expected near 22\,mag \citep{Bricman, Mallorn}. The magnitude limit of 19.5 adopted here was set by what could be reliably classified with the NTT, and lies at the bright end of the present distribution. Dedicated spectroscopic follow-up therefore remains as important as it was for this work, in which ePESSTO+ alone classified 30\% of our objects, and several
facilities are coming online to provide it: the ESO 4-m Multi-Object
Spectroscopic Telescope (4MOST), conducting the Time Domain Extragalactic
Survey \citep[TiDES;][]{Frohmaier2025}; the Son of X-shooter
\citep[SOXS;][]{Schipani2018} on the NTT; the Rubin-ESO Investigative
Spectroscopy Of Target Transient Objects (RISOTTO) survey on the ESO Very
Large Telescope (VLT); US NOIRLab telescopes such as Gemini
\citep{Mountain1998} and SOAR \citep{Krabbendam2004}; or the New Robotic Telescope (NRT) \citep{NRT}. These span a range of
apertures, and the division of labour matters: multiplexed 4-m surveys can
process large numbers at the bright end, whereas the bulk of Rubin TDEs,
peaking some three magnitudes below the limit adopted here, will require
8-m class facilities.

LSST's multi-band coverage will supply colour information, but its lower cadence relative to ZTF will limit how well colour evolution can be tracked; same-night band pairs will nonetheless provide a static colour, which is particularly valuable when the $u$ or $g$ band is available. These gaps are best filled by high-cadence photometric follow-up from the growing number of mid-sized robotic telescopes. Simple spatial and practical filtering, combined with basic photometric
discriminants, dense photometric coverage from smaller telescope networks such as BHTOM \citep{BHTOM1, BHTOM2} and rapid
spectroscopic classification matched to the depth of the target, therefore
offers a practical route to identifying SMBH-related transients in the LSST
era.

\begin{acknowledgements}
E.C., T.P., A.G. and S.B. acknowledge the financial support from the Slovenian Research and Innovation Agency (ARIS) through research programme P1-0031 and N1-0344, N1-0461, J1-2460 and J1-60014 project grants, the  infrastructural fund I0-0033 and the SNN-147362 OTKA grant of the Hungarian Research, Development and Innovation Office. Based on observations collected at the European Organisation for Astronomical Research in the Southern Hemisphere, Chile, as part of ePESSTO+ (the advanced Public ESO Spectroscopic Survey for Transient Objects Survey – PI: Inserra). ePESSTO+ observations were obtained under ESO program ID 112.25JQ. P.C. acknowledges financial support from the Secretary of Universities and Research (Government of Catalonia) and by the Horizon 2020 Research and Innovation Programme of the European Union under the Marie Sk\l{}odowska-Curie and the Beatriu de Pin\'os 2024 BP 00125 programme, and from the Centro Superior de Investigaciones Cient\'ificas (CSIC) under the Spanish program Unidad de Excelencia María de Maeztu CEX2020-001058-M, financed by MCIN/AEI/10.13039/501100011033, and by the MaX-CSIC Excellence Award MaX4-SOMMA-ICE. MN is supported by the European Research Council (ERC) under the European Union’s Horizon 2020 research and innovation programme (grant agreement No.~948381). T.-W.C. acknowledges financial support from the Yushan Fellow Program by the Ministry of Education, Taiwan (MOE-111-YSFMS-0008-001-P1), and the National Science and Technology Council, Taiwan (NSTC grant 114-2112-M-008-021-MY3).  CPG acknowledges financial support from grant RYC2024-050959-I, funded by MICIU/AEI/10.13039/501100011033 and FSE+, and from grants PID2025-172154NA-I00 and PID2023-151307NB-I00, funded by MICIU/AEI/10.13039/501100011033, FSE+, ERDF, and the EU. Additional support was provided by CSIC through PIE 20215AT016, the María de Maeztu Unit of Excellence programme (CEX2020-001058-M), and the MaX-CSIC Excellence Award (MaX4-SOMMA-ICE). LM acknowledges support through a UK Research and Innovation Future Leaders Fellowship (grant number MR/Z000076/1). T.E.M.B. is funded by Horizon Europe ERC grant no. 101125877. SN is supported by research grant (VIL60862) from VILLUM FONDEN. F.O. acknowledges support from the INAF-Large Grant 2024:"Envisioning Tomorrow: prospects and challenges for multimessenger astronomy in the era of Rubin and Einstein Telescope"; the INAF-GO Large Grant: "Exploitation of optical and near-infrared followup data of Gamma-ray Bursts" and the  INAF-MINIGRANT (2023): "SeaTiDE - Searching for Tidal Disruption Events with ZTF: the Tidal Disruption Event population in the era of wide field surveys".
\end{acknowledgements}
 
%%%%%%%%%%%%%%%%%%%%%%%%%%%%%%%%%%%%%%%%%%%%%%%%%%%%%%%%%%%%%%
\FloatBarrier
\clearpage
\bibliography{references}
\onecolumn
\begin{appendix}
\section{TDE lightcurves and spectra}

\begin{figure}[ht]
    \centering
    \includegraphics[width=0.9\textwidth,height=0.70\textheight,keepaspectratio]{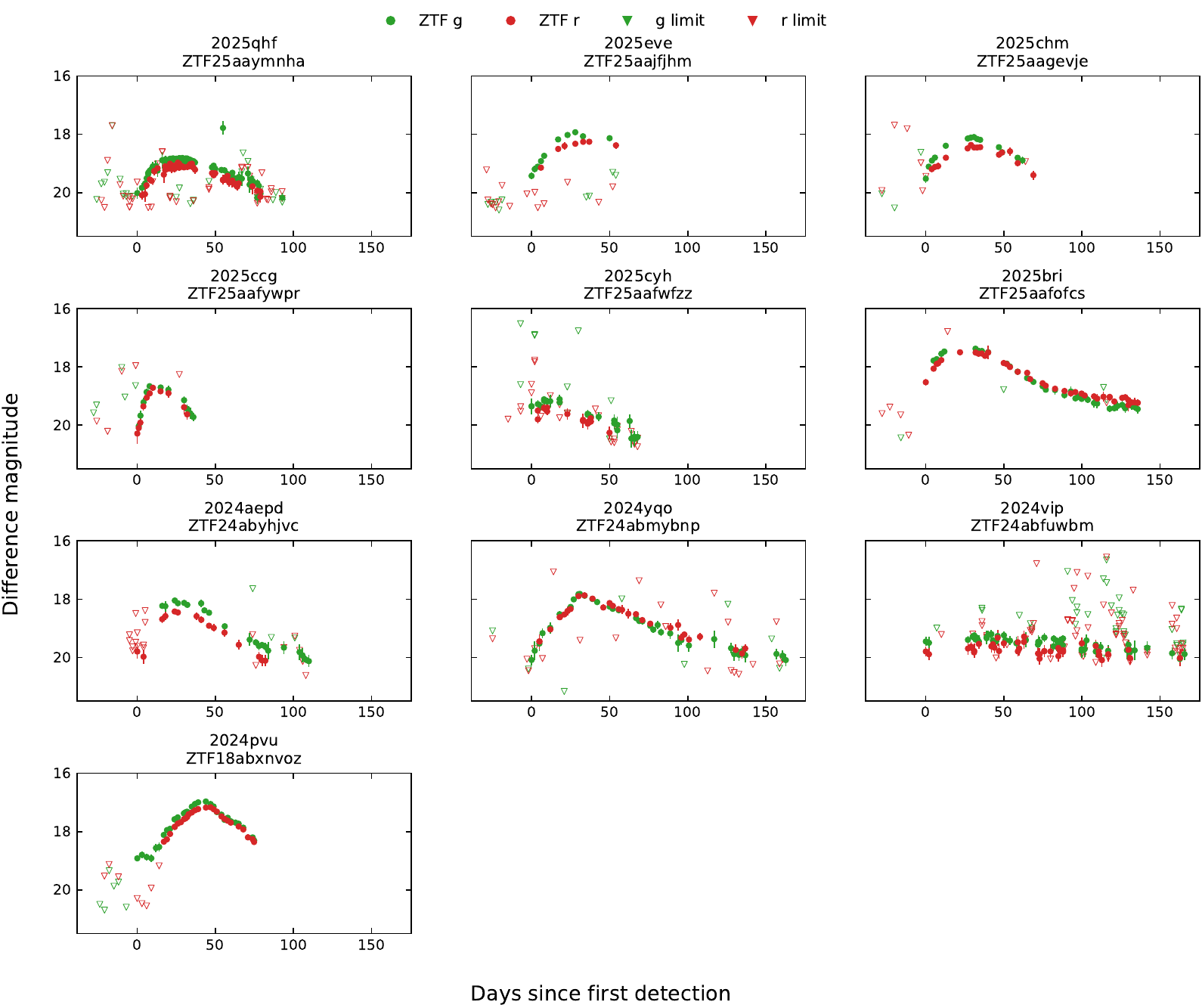}
    \caption{ZTF g and r difference photometry lightcurves of the ten selected TDEs, downloaded from the Lasair broker}
    \label{fig:TDEs}
\end{figure}

\FloatBarrier

\begin{figure*}[t]
    \centering
    \includegraphics[width=0.9\textwidth,height=0.70\textheight,keepaspectratio]{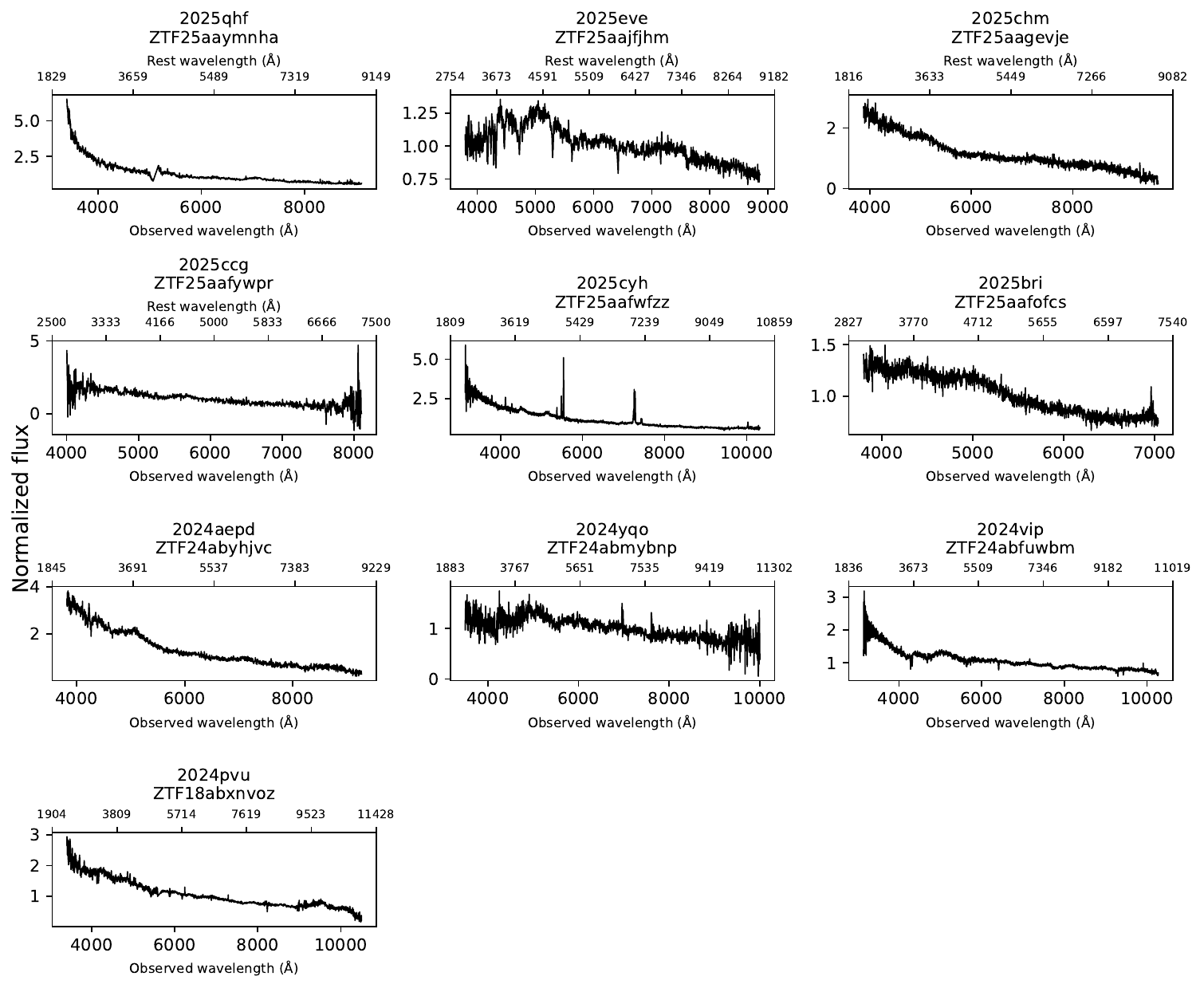}
    \caption{Classification spectra of the ten selected TDEs. The classifying groups, instruments and date of observation are shown below in table \ref{tab:TDE_spectra}}
    \label{fig:TDE_spectra}
\end{figure*}

\FloatBarrier % Forces Figure 2 to be placed before the Table

\begin{table*}
\centering
\caption{Spectroscopic follow-up summary for the ten classified TDEs recovered by our filter.}
\label{tab:TDE_spectra}
\renewcommand{\arraystretch}{1.3}
\setlength{\tabcolsep}{4pt}
\footnotesize
\begin{tabular}{l l l l l l l}
\hline\hline
ZTF name & TNS name & TDE class & Obs. date & Instrument & Telescope & Classification report \\
\hline
ZTF25aaymnha & 2025qhf & TDE-H-He & 2025-06-26 & SNIFS   & UH 88-inch               & \citet{2025qhf}  \\
ZTF25aajfjhm & 2025eve & TDE      & 2025-05-23 & ALFOSC  & NOT                      & \citet{2025eve}  \\
ZTF25aagevje & 2025chm & TDE      & 2025-02-20 & ALFOSC  & NOT                      & \citet{2025chm}  \\
ZTF25aafywpr & 2025ccg & TDE-He   & 2025-02-18 & SPRAT   & Liverpool Telescope      & \citet{2025ccg}  \\
ZTF25aafwfzz & 2025cyh & TDE      & 2025-02-15 & LRIS    & Keck                     & \citet{2025cyh}  \\
ZTF25aafofcs & 2025bri & TDE-He   & 2025-02-15 & Goodman & SOAR                     & \citet{2025bri}  \\
ZTF24abyhjvc & 2024aepd& TDE      & 2025-01-03 & ALFOSC  & NOT                      & \citet{2024aepd} \\
ZTF24abmybnp & 2024yqo & TDE      & 2024-11-09 & FLOYDS-N& Faulkes Telescope North  & \citet{2024yqo}  \\
ZTF24abfuwbm & 2024vip & TDE      & 2024-10-01 & LRIS    & Keck                     & \citet{2024vip}  \\
ZTF18abxnvoz & 2024pvu & TDE      & 2024-08-13 & DBSP    & Palomar 5.1\,m Hale      & \citet{2024pvu}  \\
\hline
\end{tabular}
\end{table*}

\FloatBarrier % Forces Figure 2 to be placed before the Table
\section{Photometric characteristics of the non-classified sample and pre-peak colour evolution of classified objects}

\begin{figure}[h!]
   \centering
   \begin{subfigure}{0.48\textwidth}
      \centering
      \includegraphics[width=\textwidth]{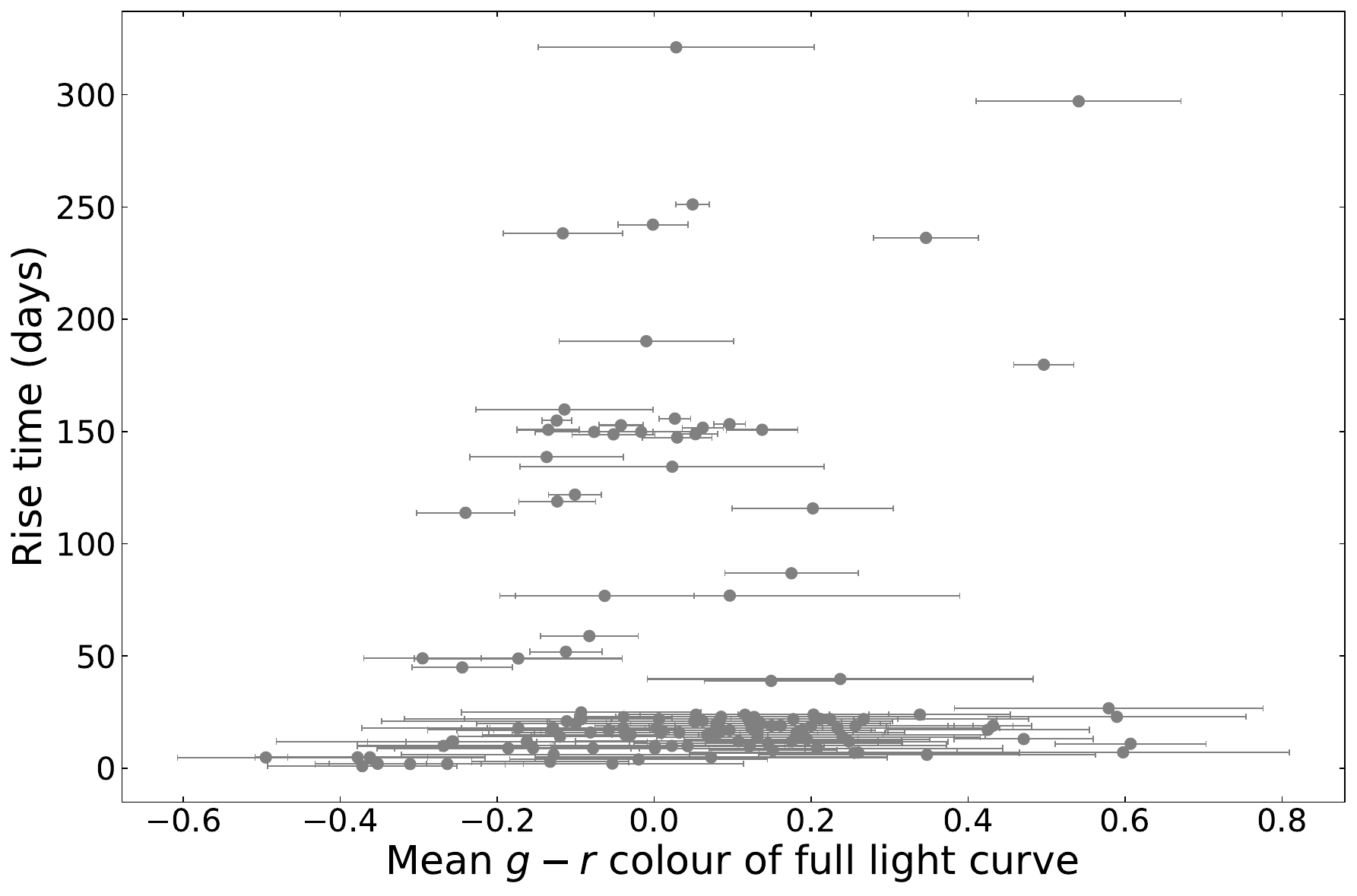}
   \end{subfigure}
   \hfill
   \begin{subfigure}{0.48\textwidth}
      \centering
      \includegraphics[width=\textwidth]{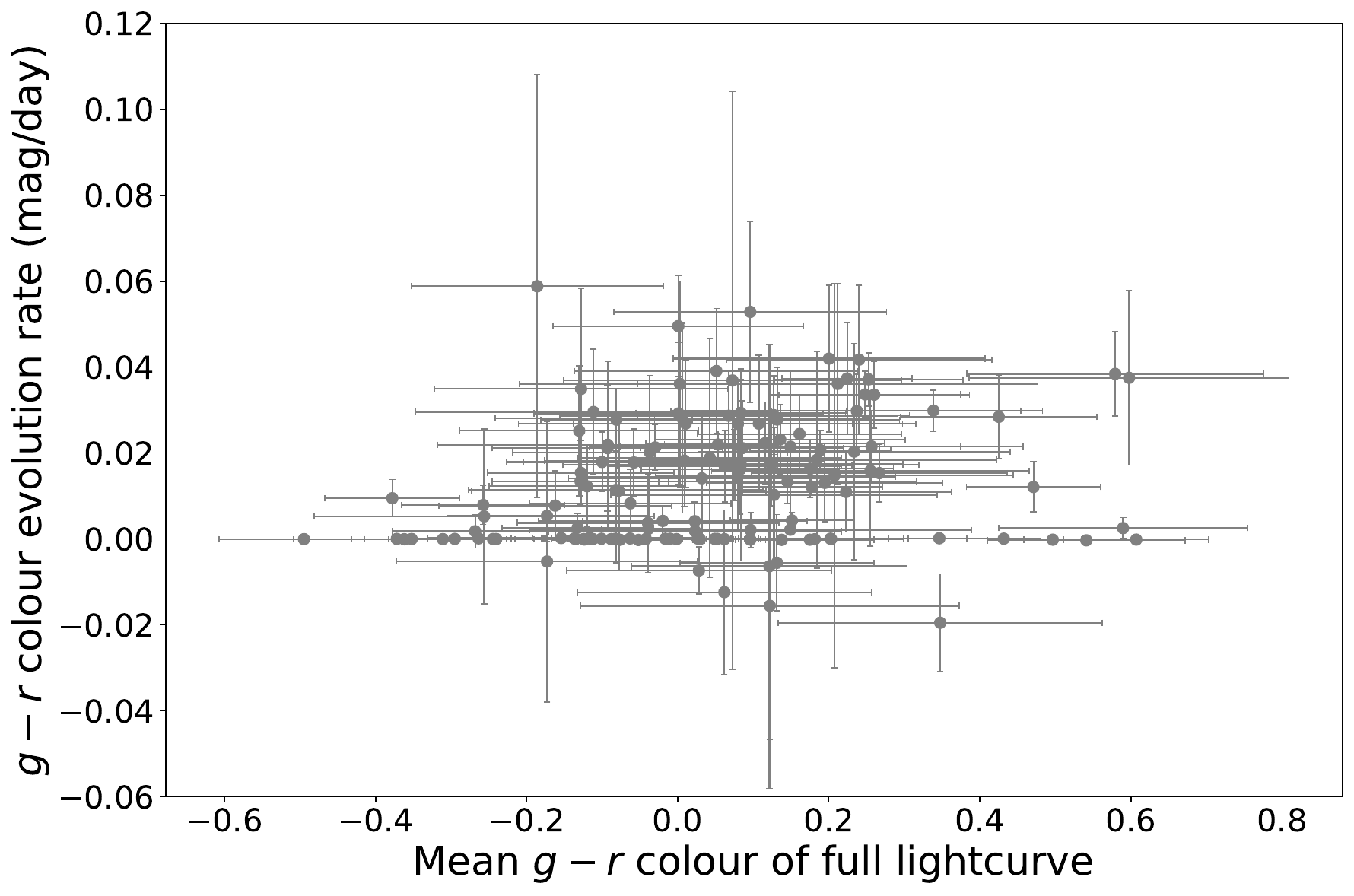}
   \end{subfigure}
   \caption{Left: Rise time against mean $g-r$ colour of the non classified sample. The distribution looks similar to that of Fig. \ref{rise_times}, with a higher number of long-rising events, which are mostly flagged as AGN by Lasair. Right: Colour evolution against mean $g-r$ colour of the non classified sample, similarly to Fig. \ref{colour evolution}. Here several objects show reddening very close to 0 (but not 0), which when looking at lightcurves seem to correspond to quasi-stochastic lightcurves like those of AGN, with no significant flare visible. Some are also flagged as potential variable stars by Lasair's contextual classifier.}
   \label{fig:not_class_analysis}
\end{figure}

\begin{figure}[h!]
    \centering
    \includegraphics[width=0.5\textwidth]{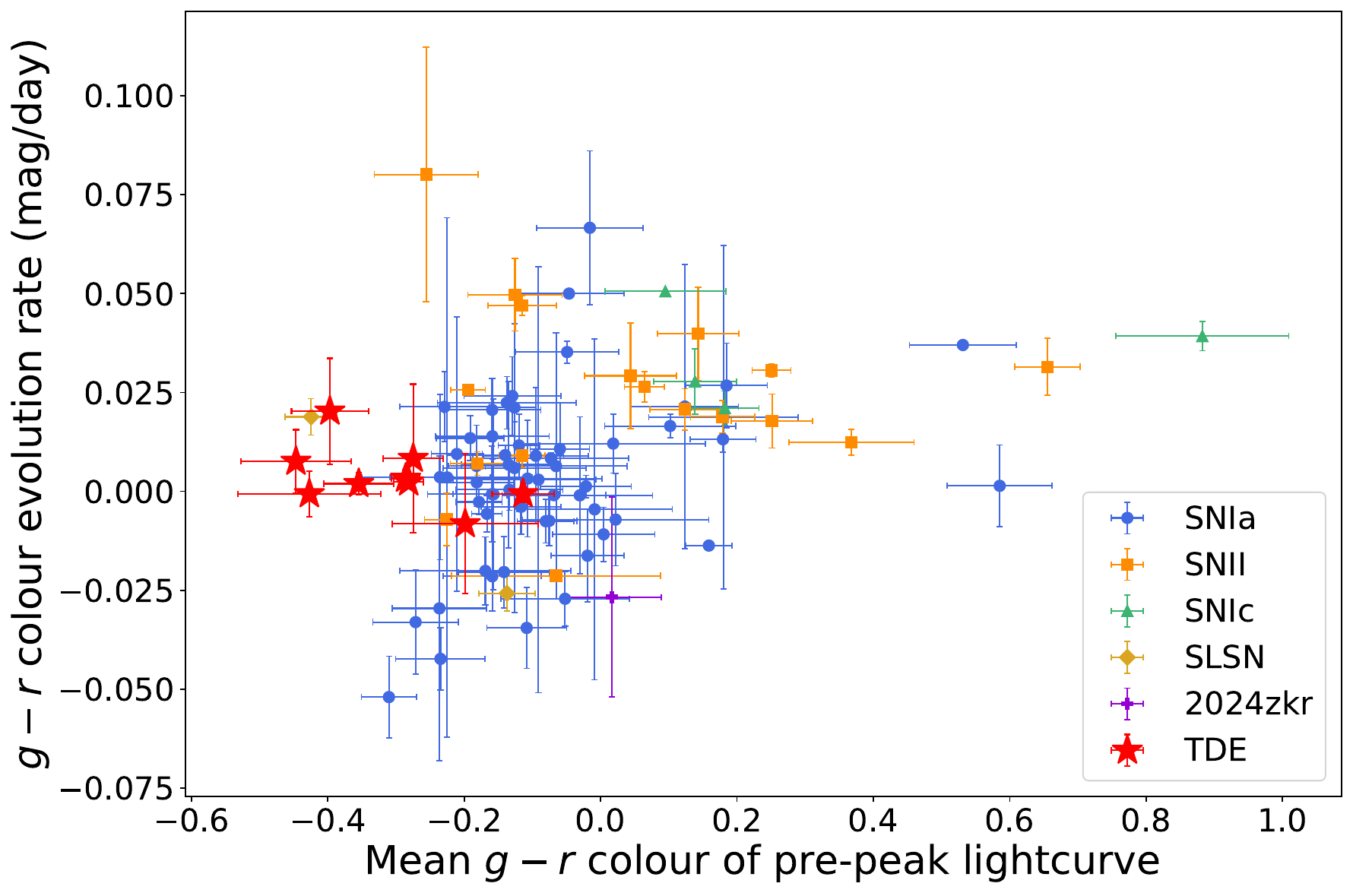} 
    \caption{Linear evolution of the $g-r$ colour, in magnitudes per day, versus the mean $g-r$ of pre-peak lightcurves, similar to Fig. \ref{colour evolution}.}
    \label{colour evolution pre}
\end{figure}

\end{appendix}

\end{document}